\documentclass[10pt,twocolumn,letterpaper]{article}

\usepackage{cvpr}
\usepackage{times}
\usepackage{epsfig}
\usepackage{graphicx}
\usepackage{amsmath}
\usepackage{amssymb}
\usepackage{url}
\usepackage{everyshi}

\usepackage[pagebackref=true,breaklinks=true,letterpaper=true,colorlinks,bookmarks=false]{hyperref}
\usepackage{breakurl}

\cvprfinalcopy 

\def\cvprPaperID{****} 

\ifcvprfinal\fi
\begin{document}

\title{Social Behavior Analysis in
Visual Human Monitoring System : \\ A Survey and Perspective}

\author{Henry Y.T. Ngan, Hideki Kawai, Kazuo Kunieda, Keiji Yamada\\
C\&C Innovation Research Laboratories, NEC Corporation\\
8916-47, Takayama, Ikoma, Nara, 630-0101 Japan\\
{\tt\small ngan.henry@gmail.com, h-kawai@ab.jp.nec.com, k-kunieda@ak.jp.nec.com, kg-yamada@cp.jp.nec.com }
\thanks{This manuscript was originally composed on March, 2009.}
}

\maketitle
\thispagestyle{empty}

\begin{abstract}
   A social behavior analysis is used to study how a group of people interacts with another group.　The analysis helps to understand how social behavior leads to its consequences such as what business decision is made after a businessmen’s meeting. In this paper, we focus on visual human motion analysis which is one important component of social behavior analysis. Human motion analysis in visual surveillance usually tracks the motion of an individual or a group of people, yet social behavior is usually neglected. This paper first delivers a literature survey of visual surveillance, with emphasis on aspects of human motion analysis and social behavior. Second, it offers a perspective for social behavior analysis in an intelligent visual human monitoring system. Third, a social interaction is induced by a social behavior between two persons, one person and one group, or two groups, hence three general scenarios of social interactions are outlined for future theoretical development. The proposed human monitoring system enables the generation of valuable quantified information of social interactions. It provides an objective approach to evaluate performance of a human organization such as a company or a school. Finally, it can raise the awareness of researchers to further explore the field of social behavior analysis.
\end{abstract}

\section{Introduction}

A social behavior \cite{Wiki09} is a behavior of a human being in the presence of other person(s) or in a society, whether or not a feedback is received. In an advanced point of view, every social interaction is induced by a social behavior. A social interaction occurs between one person and his communicated partner, as well as between one group of people and another group. Daily social interaction, such as chatting with family members, meeting with colleagues, or playing football with friends, is face-to-face and dynamic. Normally, there is a platform, namely offices, schools, and shopping malls, for social behavior to occur.

Social behavior analysis is used to study the social interaction between a group of people and another group, and its consequential effects. Social behavior \cite{McGrath} is also actively explored from the social science field, but rarely from the engineering field. A successful social behavior analysis is important and has many practical usage in the social science and business fields. Examples of the practical usage include, performance evaluation of employees at an organization based on the frequency of social interactions among themselves, and sales evaluation of service representatives at a boutique based on the frequency of social interactions between service representatives and customers.

\begin{figure}
\centering
\includegraphics[width=2.9in, height=2.8in]{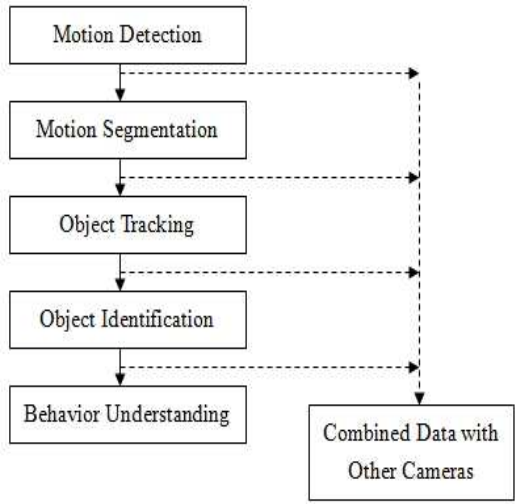}
\caption{Procedures of a typical visual surveillance. Dash arrows mean that the output of that procedure sometimes require data or analysis to be combined with other cameras.}\label{ProcedureVisSur}
\end{figure}

In this paper, human motion analysis in visual surveillance is believed to be one important component of social behavior analysis. Visual surveillance usually involves the procedures of motion detection and segmentation, object tracking and identification, and behavior understanding. There can also be a variety of data from multiple cameras.  Fig. \ref{ProcedureVisSur} illustrates the procedures of visual surveillance. An intelligent visual surveillance can assist human operators when their monitoring ability could be undermined by a vast number of cameras. It can also, perform more generic functions such as abnormal detection and alarming for security, traffic flow analysis, etc. However, previous research mainly focused on behavior understanding and interpretation on human motion data. A visual human monitoring system aims to perform, detect, track, and analyze social behaviors occurring in the visual data at an organization such as office, school or shopping mall. The ability to study social behavior in such system is usually neglected. It is worth conducting a social behavior analysis to fulfill the research gap, the details will be explained later.

In the visual human monitoring system, social behavior analysis can be carried out through video data, location history and lifelog. Meanwhile, lifelog is a set of continuously captured data of daily activities. Several areas of social behavior analysis in the system can be achieved. Examples include (1) analysis of individual human behavior in video data, (2) analysis of detected social interactions, caused by social behaviors of participants, (3) formulation of a mechanism to quantify different features in social interactions. Therefore, the later part paper will address a perspective on how to achieve such goals in the later part.
Here are three main contributions of this paper.\\

1.	A survey of human motion analysis and social behavior in visual surveillance is conducted, The research gap of visual surveillance is also discussed.\\

2.	A research perspective of social behavior analysis in visual human monitoring system is delivered. The social interaction is quantified by an intelligent visual human monitoring system. It provides additional quantitative features to evaluate an organization performance in an objective way.\\

3.	Three general scenarios of social interactions are constructed for social behavior modeling. It benefits future theoretical development of visual human monitoring system.\\

This paper is organized as follows. Section 2 describes previous research in visual surveillance including human motion analysis and social behavior. Section 3 discusses the research space of social behavior in visual surveillance. Section 4 provides a perspective of a social behavior model in human motioning system. Finally, conclusions are drawn in Section 5.

\section{Previous Research in Visual Surveillance}

Starting from late 1980s, human motion analysis \cite{Wang} attracted researchers’ concern and plenty of research were carried out in areas of human detection (low-level vision), tracking (intermediate-level vision), and recognition (high-level vision). Behavior understanding, which involves the studies of action recognition and description in human motion analysis, is usually treated as the last part at the high-level vision in a visual surveillance system. Typical approaches \cite{Wang} for behavior understanding include dynamic time warping, hidden Markov models (HMM) and neural network. Table \ref{TopicsSB} provides a summary of papers in human motion analysis and social interaction.

\begin{table}[htbp]
\setlength\tabcolsep{3.5pt}
\begin{center}
\caption{Topics Under Social Behavior.}\label{TopicsSB}
\begin{tabular}{| c | c | c |}\hline
            Reference  & H.M.A. & Social Interaction   \\\hline
			Wang et al. \cite{Wang}      & X     &       \\\hline
            Makris and Ellis \cite{Makris} & X     &       \\\hline
            Ning et al. \cite{Ning}    & X     &       \\\hline
            Ohta and Amano \cite{Ohta}  & X     &       \\\hline
            Izumi et al. \cite{Izumi}   & X     &       \\\hline
            Liu et al. \cite{Liu}      & X     &       \\\hline
            Takata et al. \cite{Takata}  & X     &       \\\hline
            Doherty and Smeaton \cite{Doherty}  & X     &       \\\hline
            Singletary and Starney \cite{Singletary} &    &    X  \\\hline
            Hoey and Little \cite{Hoey} & X    &    X   \\\hline
\end{tabular}
\end{center}
Remark: H.M.A.= Human Motion Analysis
\end{table}

\subsection{Human Motion Analysis}

Some recent work \cite{Makris,Ning,Ohta,Izumi,Liu,Takata,Doherty} has been done on human motion analysis. The first example is an activity-based semantic scene model of a video surveillance system, developed by Makris and Ellis \cite{Makris}, for tracking specific areas such as entry/exit zones, stop zones, etc. Their model emphasizes on object motion tracking to follow its trajectories, equivalent to the location history, on various scene locations. An adaptive pixel-wise model for a sequence of video frames is defined as a Gaussian mixture model (GMM) for the background. Based on GMM, a topographical representation of the scene elements as nodes is used as a basis of Bayesian belief network (BBN) for learning. Afterwards, a HMM theory for activity analysis is utilized to overlay the network. Some advantages are listed. First, their method for learning the HMM parameters is in a relatively simplified way. Second, their unsupervised approach allows the system to observe and learn the environment. Third, the method is tested from individual camera to multi-cameras situation so that it can connect the paths between cameras. However, in evaluation, only 85.7\% accuracy is achieved to detect the entry/exit zones from 70 experiments. Although they performed a second test on 100 previous unseen trajectories with manual labels on routes and claim to have reached 97\% accuracy, not much details about this test is provided.

The second sample is a set of spatial-temporal words, defined by Ning et al. \cite{Ning}, from unlabelled data representing various human behaviors. It is used for searching human behaviors in videos by unsupervised learning. A patch-based feature is obtained by the histogram of responses of a bank of 3D Gabor filters and followed by a MAX-like operation. Then, the human behavior similarity is estimated by the discrepancy of spatial-temporal words frequency. Searching is accomplished by a correlation of the similarity between the query video and the search video at all sliding windows. The advantages of their proposed method are that, (1) the spatial-temporal words not only captures the intrinsic information of motion and appearance of human behaviors, but also speeds up the scanning through integral histograms, and (2) its patch-based feature is locally invariant to a range of scale and position variations. However, the performance evaluation is only carried out by 2 videos data of 31 frames and 400 frames, which is not convincing enough.

Ohta and Amano \cite{Ohta} constructed a moving mobile robot to recognize human behavior of the elderly. The human behavior is described as a time sequence of posture, for which 3 categories are considered: stop, moving forward and moving backward. Their study firstly captures human body by a stereo camera, and secondly generates three-dimensional (3D) point data after stereo vision processing. Those 3D point data would be transferred into 3 two-dimensional (2D) binary images, in order to determine the hand position, and the distance between the human and the robot. This distance is constrained to the human hand being captured by the robot. Although they showed that the robot can correctly measure the hand position, their experimental set-up is limited by assumptions of human position, reaction range and speed.

Another two examples use a fuzzy based approach. The first one is from Izumi et al. \cite{Izumi}. The human posture image is firstly captured by a binocular camera. Utilizing background difference, a moving human body and its position information is calculated by a stereo method. Three human postures, namely standing, crouching and lying, are later to be determined using an aspect ratio of the human region and the lateral deviation of the upper and lower half-of-body regions of the detected human. Lastly, human behavior is predicted by the change in the time-series and performed by a fuzzy neural network (FNN). The input features of their FNN are characterized by the human velocity, the angle of the human relative to an object, and the distance between the human and an object. Though the method is shown with figures, which successfully estimate the human behavior, only 1 scenario and 1 human motion with 20 consecutive segmented images are evaluated. Different varieties of input features are not investigated so the generalization of their approach is still questionable.

The second one from Liu et al. \cite{Liu} delivered a fuzzy Petri nets-based method to a verification and validation technique in a fuzzy rules-based human behavior models for military simulations. The method includes verification of fuzzy rule bases, static validation of human behavior models, and dynamic validation of human behavior models. Herein, a formal description with theoretical foundation for human behavior models is provided, yet there is no experimental result to allow the reader to understand their real performance.

The last two examples employ the facilities of lifelog. A lifelog is not only composed of visual data, but also with other media data such as audio, body movements and positions. Takata et al. \cite{Takata} have recently analyzed an individual’s daily activities through lifelog, acquired by different wearable sensors based on an action-oriented model and a space-oriented model. These wearable sensors consist of camera, accelerator, heart rate sensor, global positioning system (GPS) signal receiver and computer. An integrated technique is applied to process the lifelog data using correlation between different types of captured data from multiple sensors. The structured lifelog images of body motion data, biological data and location information are claimed to be better than vision-based technology. Here are 2 limitations in their models. First, there is a deficiency for huge object moving in lifelog images segmentation. Second, only 4 different application cases are tested.

Doherty and Smeaton \cite{Doherty} proposed a passively capturing wearable camera, SenseCam, to acquire visual lifelog in order to identify important events. The camera produces a personal lifelog and useful information as a human memory aid. Their model is capable of automatic emphasis on relative important events and elimination of the routine events. A concept of novelty, defined as rarely occurring event, is utilized to determine which event is more unique than other events. A total of 288, 479 lifelog images of 3,445 events from 6 users over 1 month are evaluated for the best novelty approach, the best face/conversation detection approach, and the combined face detection and novelty approach. The last approach is found to be superior to other approaches. The questionable part of their research is that the user of SenseCam is the only person who judges various experiments on his own database.

\subsection{Social Behavior}

\begin{figure}
\centering
\includegraphics[width=2.8in, height=1.6in]{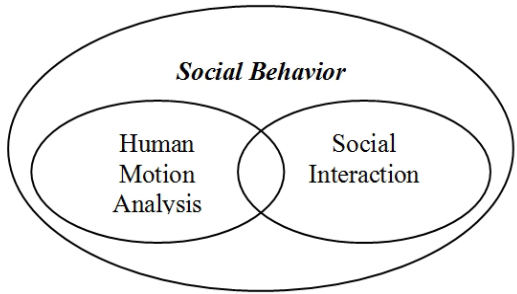}
\caption{In visual surveillance, the research area of social behavior includes human motion analysis and social interaction. }\label{SBanalysis}
\end{figure}

In visual surveillance, social behavior is relatively new and rare in the literature and includes the research of human motion analysis and social interaction (Fig. \ref{SBanalysis} ). Only a few papers have attempted to study this area. Two examples \cite{Singletary, Hoey} are described below.

An early related work, by Singletary and Starney \cite{Singletary}, utilized a face detector of a wearable system to assist with social engagement. Their system performed face recognition and recognition once a social engagement happened. Over 300 participants in a conference are captured one or more time over a 10-hour video recording used for analysis. Afterwards, over 90\% detection rate is achieved for the user starting to interact with other individual(s).  They collected video data in an open area with direct sunlight and a darkened conference hall, both of which were highly unconstrained. A three-state HMM is applied to visual model of human interaction. There are three limitations. Firstly, it only focuses on the interaction of 1 person and another one, without other kinds of interaction. Secondly, without observer mode support, the user is required to use the system directly. Thirdly, no other human motions such as gestures are captured for analysis.

A recent good example is from Hoey and Little \cite{Hoey}, they tried to discuss social behavior in their research. They analyzed human motion from a learning decision theoretic model, called partially observable Markov decision process (POMDP) from video data. In particular, a dynamic Bayesian network integrates the observed video into the POMDP for supplying spatial-temporal abstractions amenable to decision making at the high level. Their model can automatically learn the distinguishable behaviors and discover its categories through a training process. It is evaluated by 3 kinds of human interactions through facial expressions including a single player imitation game, a gestural robotic control problem, and a card game played by two persons. There are 3 limitations in their method: (1) insufficient validation due to simulations solely on selected actions and gathering rewards online by simple prediction actions from a human, (2) applying fully observed state space without concern of unobserved variables, and (3) using a small amount of displays and only allowing merging states during learning. Moreover, their method did not consider interactions among various groups of people.

\section{Research Space of Social Behavior Analysis in Visual Surveillance}
From the literature survey above, not much research has been studied for social behavior, besides human behavior understanding. In addition, the research of social behavior is still at its early stage. Here are some shortcomings of the current models \cite{Singletary, Hoey} in social behavior analysis in visual surveillance. First, only a simple case of social interaction between one individual and another one is modeled. Second, only part of a human body, such as face, is being focused in visual detection and tracking. Third, a small amount of samples are evaluated in the previous methods. An ideal social behavior study of human motion analysis should possess two characteristics, (1) generality to analyze different scenarios of social interactions among people and take different parts of human body into account for social behavior analysis, (2) reliability to apply to lots of scenario samples of social interactions.

   Here are three research motivations to construct a social behavior model of human motion analysis.\\
   
1.	There is no existing visual surveillance model, which enables the analysis of social behavior among people. Hence, an intelligent human monitoring system should be established.\\

2.	Social interactions contain numerous and diverse human behaviors. An understanding of a valid social behavior in visual data can provide more in-depth information of advanced human behavior.\\

3.	The output of social behavior analysis from human monitoring system can be generated as quantified information, which is useful for further study in organization performance.
Therefore, we would like to propose a perspective on a model of social behavior analysis of human monitoring system in the following section.\\

\section{Perspective of a Model of Social Behavior Analysis of Human Monitoring System}

\begin{figure*}[tH]
\centering
\includegraphics[width=5.6in, height=2.6in]{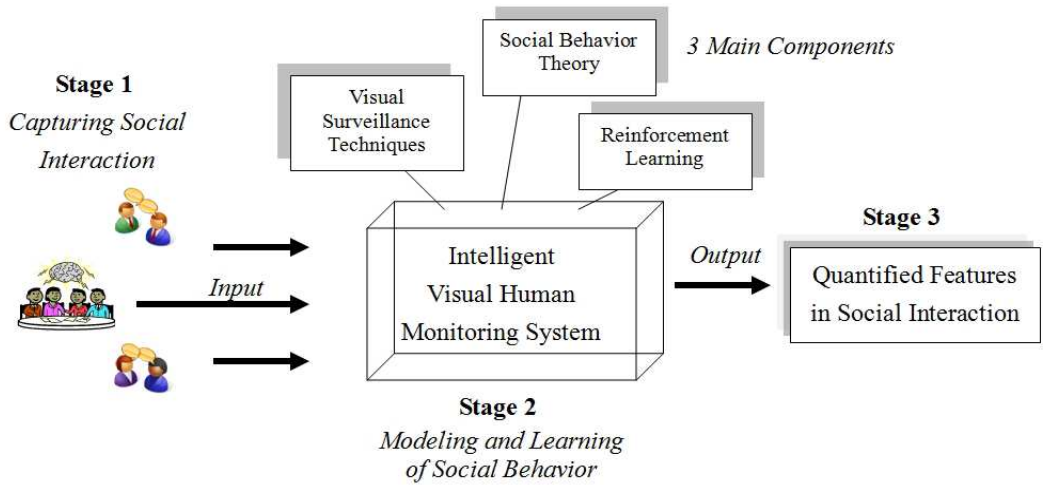}
\caption{Conceptual model of social behavior analysis in human monitoring system.}\label{ConceptualModel}
\end{figure*}

\subsection{Main Stages of The Model}
A conceptual model of social behavior in a human monitoring system is presented in Fig. \ref{ConceptualModel}. The model consists of 3 main stages:
(1) Input of visual data of social interaction,
For a certain period of time, a social interaction scene should be captured, detected and segmented, as the input for the visual human monitoring system.
(2) Modeling and learning in intelligent visual human monitoring system,
 Stages of learning and classification are involved here. A visual human monitoring system first builds models of social interaction for learning by setting up camera(s) on some locations, and starts to perform classification for detected social interactions in the second step.
(3) Providing quantified features of social interaction.
 The intelligent system should be able to provide quantitative features after classification of various social interactions. It helps the researchers to employ these features in the coming quantitative analysis such as performance evaluation for employees in future research.

\subsection{Tasks of The Proposed Model}
Essential tasks of the proposed model in the near future are outlined as below.\\

1.	Definitions of Various Social Behaviors. Social behaviors deserve to be well defined in acquired visual human motion data. Gestures with various parts of human body such as face, hands and legs should be taken into account for motion analysis.\\

\begin{figure*}[tH]
\centering
\includegraphics[width=6in, height=2.2in]{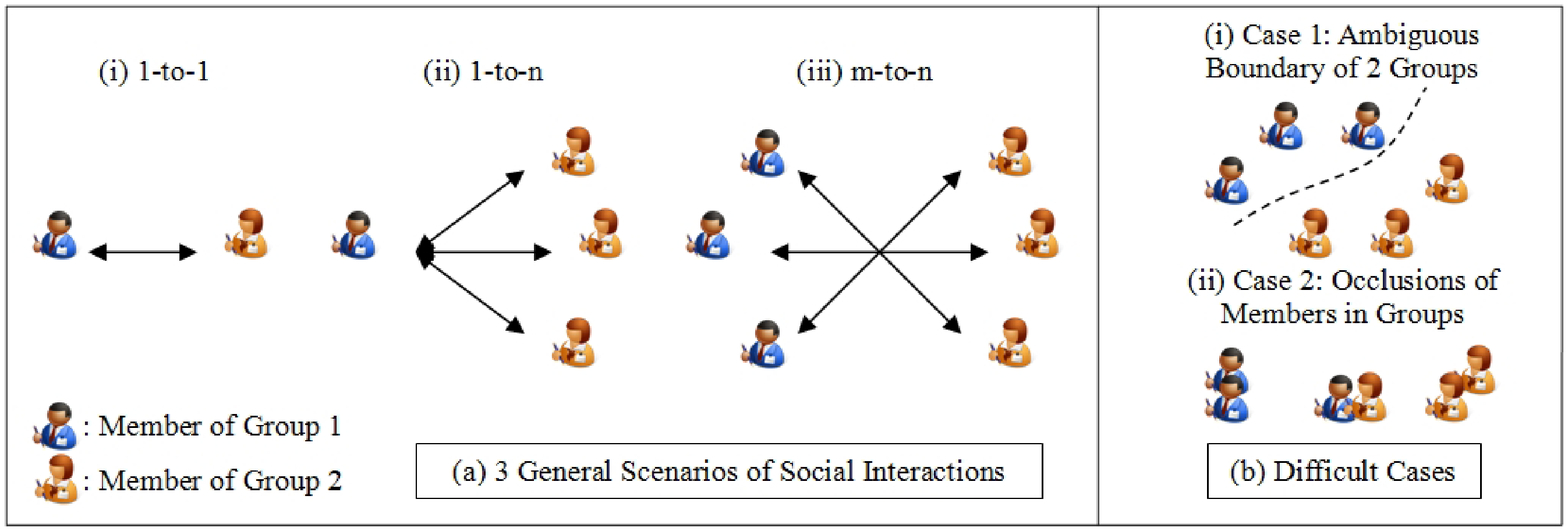}
\caption{(a) Three general scenarios of social interactions (i) 1-to-1, (ii) 1-to-n, and (iii) m-to-n, (b) two difficult cases, (i) case 1 of ambiguous boundary of 2 groups, and (ii) case 2 of occlusions of members in groups. }\label{SIscenarios}
\end{figure*}

2.	Scenarios Modeling. Social interaction is in a dynamic and complex form, hence some scenarios are necessarily established for modeling. As a valid social interaction involves 2 groups of people, 3 general scenarios can be constructed mathematically. Simply, they are the interactions of (a) 1 person with 1 person (1-to-1 in Fig. 4(a)(i)), 1 person with 1 group (1-to-n in Fig. \ref{SIscenarios}(a)(ii), where n is an integer), and 1 group of people with another group (m-to-n in Fig. \ref{SIscenarios}(a)(iii), where m, n are integers).  Two examples of complex scenarios are ambiguous boundary for classifying interaction groups (Fig. \ref{SIscenarios}(b)(i)), and occlusion between members in groups (Fig. \ref{SIscenarios}(b)(ii)).\\

3. Constraints.  There will be many constraints at the social interactions in captured visual data that should be tackled. Examples include (1) similar gestures among similar social behaviors during chats and meetings and (2) similar textures between people and background.  They should be tackled by pattern recognition techniques.\\

4. Collection of Database. A collection of database should be created for various social interactions exhibited at those scenarios.\\

5. Generic Approach. In our system, a generic approach is necessary for design and installation to detect, recognize and classify different categories of social behavior in visual data. It should be composed of 3 main components, shown in Fig. \ref{ConceptualModel}, such as visual surveillance techniques, social behavior theories and reinforcement learning.

\subsection{Applications of The Proposed Model}

After a model of social behavior analysis in human monitoring system is constructed, it can be further applied to performance evaluation of organizations such as business offices, classrooms and conference rooms. Organization performance \cite{OrgPer-Wiki} is defined as the actual output or results of an organization to be estimated against its intended outputs, such as goals and objectives.

A usual way for current measurement metrics for performance in an organization includes traditional surveys and statistics. With surveys, a measurement indicator \cite{Trea} is built from all narrative, qualitative and texture information at the U.S. IT Department of the Treasury for their performance evaluation of IT investment. However, this kind of measurement is qualitative and can easily be biased from different sources of focus groups. With statistics, there are 2 measures. First, typical result-based measures \cite{Meyer} in the commercial world utilize revenues, gross margins, cost of goods sold, capital, assets, and debts. Second, process-based measures \cite{Meyer} use the percentage of unique parts in time, cost, quality and product performance. However, these metrics are just financial indicators and based on apparent facts. In contrast, there is no quantitative way to judge how employees interact.

\begin{figure}
\centering
\includegraphics[width=3.2in, height=1.5in]{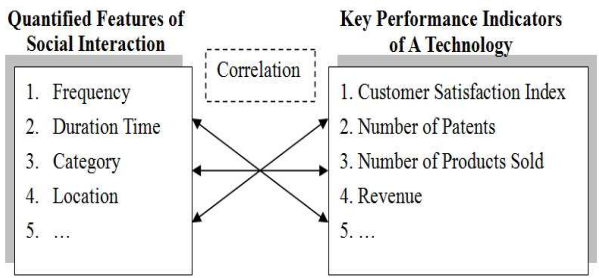}
\caption{A correlation study between quantified features of social interactions and key performance indicators of a technology corporation for performance evaluation.}\label{CorrelationStudy}
\end{figure}

By the quantified features of social behavior analysis, a new objective approach is proposed to evaluate the performance of an organization. An example is shown in Fig. \ref{CorrelationStudy}. A correlation analysis can be performed by the quantified features of social interactions obtained by visual human monitoring system and features of a business company.

Here are some quantified features which can be obtained from the visual human monitoring system, shown in the left hand side of Fig. \ref{SIscenarios}, such as
(1) frequency of social interactions (e.g. times per week),
(2) duration time for each interaction (e.g. length in seconds, minutes or hours),
(3) category of each detected interaction (e.g. 1-to-1, 1-to-n or m-to-n interactions),
(4) location of social interaction happened (e.g., meeting room, corridor, entrance, exit), and
(5) type of people involved in one social interaction (e.g. employees or customers).
In addition, an example of some key performance indicators of a technology corporation for the evaluation is shown in the right hand side of Fig. 4, including
(1) customer satisfaction index,
(2) number of patents,
(3) number of products sold, and
(4) revenue, etc.
All of them should be calculated within a certain period of time in order to compare with quantified features of social interactions before.
	Some potential contributions for this correlation study consist of the understandings of (1) the relationship between social interaction and monthly revenue, (2) the relationship between social interaction and patents, (3) the relationship between employee performance and office layout design for social interaction, (4) the relationship between team-based performance and individual performance for the employees, and (5) how to improve the interaction between employees and customers, etc.

\section{Conclusions}
This paper provides a survey of human motion analysis and social behavior in visual surveillance. A perspective of conceptual model of social behavior analysis of a human monitoring system is provided. Three general scenarios of social interactions are modeled as well. The system aims at exploring quantified information among visual social behavior of people and generating valuable metrics for further evaluation of performance in human organization. It contributes by using an objective way to evaluate the performance of an organization.



\end{document}